# Small Data Generative AI for Culturally Specific Artistic Creation: A Case Study on the Sonic Space of *sarīr*

**Arshia Sobhan, Paul Triana, Philippe Pasquier**
Simon Fraser University
Vancouver, Canada
asobhans@sfu.ca, paul_triana@sfu.ca, pasquier@sfu.ca

**Abstract**

Large-scale generative AI (GenAI) models have become increasingly integrated into contemporary creative practice, yet they often reproduce cultural biases and fail to meaningfully engage with non-Western artistic traditions. In response, this paper explores a small-data approach that foregrounds artist-curated and culturally informed dataset creation and model training. Through a case study on *sarīr*—the distinctive sound produced by the reed pen in Persian calligraphy—we demonstrate how small-data GenAI can operate within a highly situated cultural context that lies beyond the representational reach of large-scale GenAI models. We construct a bespoke *sarīr* audio dataset and train two models, RAVE and StyleGAN2, adapting each to the sonic domain through audio synthesis and spectrogram generation. The resulting models learn the fine-grained timbral and gestural nuances of *sarīr*, enabling both real-time interaction and extended generative output. By comparing these results with state-of-the-art text-to-audio models, we clearly demonstrate the disparity in their performance and highlight the expressive possibilities unlocked through the small data approach. This focused case study demonstrates the feasibility of small-data GenAI for culturally specific artistic creation and underscores the value of further investigations that extend this approach to other cultural materials and artistic domains.[1]

## Introduction

In recent years, generative artificial intelligence (GenAI) has become deeply integrated into artistic creation. Users now work with GenAI models that can synthesize text, image, sound, music, and video with minimal technical intervention, extending creative practice across multiple modalities [1]. The widespread availability of prompt-based interfaces has made these models accessible to a broader public.

While these models have lowered the barrier to entry for AI-assisted creation, they also have prompted significant critique from artistic and research communities. These concerns span a sense of diminished authenticity and authorship [2], a reduced capacity for surprise [3], the pressure to translate tacit artistic intentions into explicit prompts [3, 4], aesthetic homogenization [4], limited creative control [5], and wider concerns related to sustainability and data-scraping ethics [6].

A growing body of research has also documented the cultural biases, misrepresentations, and exclusions present in such models. Text-to-image models, in particular, have been shown to reproduce stereotypical and often harmful representations of non-Western cultures (e.g. [7–11]). These issues are not confined to the visual domain and also manifest in large-scale GenAI models for other modalities, such as text [12] and music [13].

The presence of such high-level cultural issues in large-scale GenAI (such as bias and misrepresentation in the depiction of people, places, or cultural artifacts) suggests that the problems become even more pronounced when it comes to culturally specific art forms, since they carry layers of tacit knowledge, aesthetic conventions, and embodied practices that are deeply situated within specific cultural histories. However, there are only a few studies (e.g. [10]) that focus on the subtleties of specific non-Western art forms to investigate how they are treated by large-scale GenAI.

In response to these issues and limitations, recent research has begun to advocate for a small-data approach [14–16]. This approach focuses on artist-created (or -curated) datasets that represent a specific aesthetic domain and enable direct control over the material and iterative control over the generative process, rather than relying on massive, internet-scale opaque datasets. This approach relies on training generative models exclusively on these bespoke datasets, rather than adapting large-scale pre-trained GenAI models.

We propose the small-data approach as a method that enables culturally specific artistic creation by allowing artists to treat data as expressive material and make aesthetic and curatorial decisions that embed tacit cultural knowledge into the process. In this approach, the artistic labour is invested in creating or collecting, selecting, and shaping the dataset, as well as iteratively training and evaluating the model. This labour becomes an act of cultural alignment, where artistic decision-making intertwines with the interpretive work of representing a cultural form. We explore how this process can create space for situated and embodied artistic knowledge, allowing cultural understanding to be encoded through the careful design of small, specific datasets. We believe this approach helps counteract the narrow cultural coverage and aesthetic flattening of large-scale GenAI models while restoring artistic control and cultural authenticity to the creative process.

[1] Samples from the dataset and the results of our case study are available on: `https://www.metacreation.net/sarir`

We investigate this approach through a focused case study situated in Persian calligraphy. This art form is rooted in the Arabic script, which is a significant cultural element across many regions and has given rise to diverse calligraphic traditions, including the distinct Persian styles practiced in Iran [17, Pt. IV]. Our case presents an exploration into the creative potential of *sarīr*, which is the distinctive sound of the calligraphy reed pen as it is drawn across paper. This sound is a highly specific cultural element that is significant and familiar within its cultural context, yet it remains underrepresented as a peripheral modality of Persian (and broadly, Arabic) calligraphy. By engaging with such a specific and unique cultural element, this study demonstrates how small-data GenAI can operate meaningfully within highly situated artistic contexts, which lie far beyond the representational reach of large-scale GenAI models.

To carry out this exploration, we create a custom *sarīr* dataset and train small-data models using two distinct GenAI architectures: one specifically designed for audio, while the other is mainly for visual domains that we adapt creatively to extend the expressive possibilities. We present the outcomes of our experiments and incorporate them into a series of audio-visual sketches that illustrate the creative potential of the trained models. We also compare these results with outputs from state-of-the-art large-scale text-to-audio models. Finally, we discuss the challenges and limitations of this approach and outline its broader implications.

We position our investigation within a practice-based research framework [18], which emphasizes the integration of tacit and embodied cultural knowledge into the process of creation [19]. Within this framework, our investigation is shaped by an ongoing negotiation between technical affordances and cultural-aesthetic considerations in a continual reciprocal exchange as the work unfolds.

This study also foregrounds the role of cultural agency and authorship in practice-based research. The dataset creation process, as well as the project's curatorial and aesthetic decisions, draws on the first author's lived experience within Iranian culture and formal training in traditional Persian calligraphy. This is in addition to his work with GenAI for artistic creation and familiarity with its technical dimensions. The second author contributes expertise in audio synthesis and GenAI, supporting the technical development of the work, while the third author provides supervision and guidance for this research. Through this collaboration, we aim to keep the exploration culturally informed while engaging with the technical aspects of using GenAI.

Lastly, we acknowledge that cultural production is inherently diverse and resists generalization. As such, our aim is not to offer a generalizable framework, but to show how a small data approach can serve as a promising alternative for artists working with culturally specific art forms. We hope this study provides motivation and methodological insight for artists and creative practitioners exploring culturally grounded approaches to GenAI.

# Background

## The Sound of *sarīr*

Although Persian calligraphy is primarily presented and studied as a visual art form, within its cultural context, it constitutes a multisensory experience that extends beyond sight [20]. Among its lesser explored dimensions is its auditory character—the distinctive sound produced by the reed pen as it moves across the paper. This sound, resulting from the friction of the reed through ink on the surface of paper, has a signature tonal quality: a fine creaking resonance that intensifies along the elongated gestures of the written strokes.

This sound, while not a primary modality of calligraphic practice, has long been acknowledged and appreciated within Iranian culture. It is likewise publicly familiar and commonly heard whenever calligraphy is practiced, often carrying a positive connotation and perceived as a meditative sound. In Farsi (the language spoken in Iran), *sarīr* (صریر) carries a poetic meaning, referring specifically to "the music" or "the singing" of the reed pen on paper while writing, as described in the Mo'in Encyclopedic Dictionary [21, p. 988]. This interpretation reflects a cultural sensibility that perceives calligraphy not only as inscription but also as a performance of rhythmic and sonic events.

Auditory dimensions of calligraphic practice also have deep connections in Persian literature and mysticism. A particularly resonant example appears at the opening of the *Masnavi* [22] by the 13th-century Persian poet and mystic *Rumi* (1207–1273). The exordium ("The Song of the Reed") begins with:

> *Listen to the reed how it tells a tale, complaining of separations—*
> *Saying, "Ever since I was parted from the reed-bed, my lament hath caused man and woman to moan."* [22, p. 5]

While the reed in this passage traditionally refers to the "reed flute," it draws parallels with the "reed pen," since both are made from the same material (*ney*) and share an internal hollow. The sonic characteristics of the reed pen (creaking and screeching as if moaning) further strengthen this parallel, giving *sarīr* a symbolic affinity with the lament of the reed flute [23, p. 120].

From the perspective of acoustic ecology, *sarīr* can be understood as a soundmark: a community-specific sound that is uniquely recognized and valued within its cultural setting [24]. In this sense, it is part of the acoustic identity of its cultural context and what makes the auditory life of the community unique.

Although its cultural significance and its abstract, evocative sonic character invite artistic exploration, to our knowledge, no dedicated dataset or systematic study of its sonic dimension currently exists, and there are few examples of *sarīr* being used as creative material. One notable example is a live audio-visual performance (involving the first author) in which *sarīr* becomes the primary sonic source for generating a real-time soundscape that resonates with the visual gestures of calligraphy [25, Sec. Reprising Elements], which shows its creative potential.

### Small Data GenAI

Although small-data approaches predate the rise of large-scale AI, they have recently gained momentum as a mitigation strategy for the limitations of large-scale GenAI, particularly in creative contexts where personal and culturally situated expression is essential. Instead of training models on massive, general-purpose datasets, small-data methodologies rely on limited, curated datasets that allow for intentionality, control, and personal expression in GenAI practices. In creative contexts, this mindset repositions data as creative material and a site for aesthetic and conceptual intervention. Vigliensoni et al. [14] argue that small datasets are often more compatible with the goals of artistic creation, since artists typically seek distinctive, situated, and experimental outcomes rather than generalized outputs that appeal to all users. They also argue that working with smaller corpora lets artists iteratively adjust datasets and architectures, enabling a more interactive and reflective relationship between maker and model.

Small-data practices have also been promoted as a means of personalization and ethical engagement in GenAI. Abuzuraiq and Pasquier [16] emphasize that training or fine-tuning models on small, artist-created datasets can enhance authorship and agency. Similarly, Bryan-Kinns et al. [15] highlight that small datasets support responsible and low-resource AI development, reducing both environmental impact and cultural bias by decentralizing control away from large corporate infrastructures. Similarly, Sobhan et al. [25] highlight the benefits of accessible and local small-data tools to respond to artists' concerns about putting their data online.

The small-data mindset has also led to efforts to make such approaches more accessible to artists without requiring advanced technical skills. Platforms such as Playform [26] and Autolume [25] offer artist-oriented GenAI environments for visual domains. While Playform provides a paid online platform with a simplified interface for training and exploring models, Autolume promotes locally run workflows and offers expanded features suited for real-time and interactive settings.

### Large-Scale Text-to-Audio GenAI Models

Recent advances in text-to-audio generation have produced large-scale models trained on diverse audio collections, enabling the synthesis of sound effects, music, foley, environmental ambiences, speech, and other audio textures from textual prompts [27]. Notable open-source options among these models are AudioLDM2 [28], Stable Audio Open [29], and AudioGen [30]. These models are trained on large-scale datasets that introduce cultural and geographic biases through reliance on YouTube-derived data [31–33] and, in some cases, ChatGPT-assisted annotations that reflect cultural assumptions [34]. These datasets also raise copyright concerns due to ambiguous licensing, with Stable Audio Open as an exception through its use of openly licensed material.

Technical constraints further shape these systems: AudioGen and AudioLDM2 generate lower-fidelity 16 kHz audio, with AudioLDM2 limited to short outputs, while Stable Audio Open supports higher-fidelity 44 kHz audio and longer generations. Although only Stable Audio Open acknowledges reduced performance with non-English prompts, all three models are trained on English and Western-biased datasets.

## Case Study

Our case study explores the creative possibilities that emerge when small-data GenAI is applied to *sarīr*. In the following sections, we outline our process of constructing a bespoke dataset, training two GenAI models that accommodate small datasets, evaluating their generative behaviours in relation to results from state-of-the-art text-to-audio models, and the artistic sketches that emerge from this exploration. Samples from the *sarīr* dataset and generated outputs are available on our webpage: `https://www.metacreation.net/sarir`

### Results from Text-to-Audio Models

Before training our own models, we first evaluate how state-of-the-art text-to-audio models respond to *sarīr* in order to establish a baseline for comparison. We select three widely used, high-performing open-source text-to-audio models (AudioLDM2, Stable Audio Open, and AudioGen), which we introduced earlier in the Background section, and test them using a range of prompts. These included prompts such as "Persian calligraphy *sarīr*" and "the sound of a reed pen writing in Persian calligraphy," as well as more descriptive variants such as "the creaking sound of a reed pen during Persian calligraphy writing." To broaden the prompt space, we also generated outputs using "Arabic calligraphy" in place of "Persian calligraphy."

Across all three models, the results show no meaningful connection to the sonic domain of *sarīr*. The generated audio samples span unrelated categories (gibberish speech and singing, musical textures, mechanical noise, typewriter sound, and other various forms of abstract sounds), none of which bear resemblance to the subtle creaking resonance or gestural rhythm characteristic of *sarīr*. In a few cases, the outputs contain faint traces reminiscent of a pen or pencil moving quickly across paper, but these lack the timbral characteristics, pressure-based modulation, and slow, elongated gestures of the reed pen that define the calligraphic soundscape.

These observations demonstrate a complete failure of these text-to-audio models to represent this culturally specific sound. The results reinforce our broader argument that certain non-Western artistic materials fall entirely outside the representational reach of large-scale models trained on internet-scale datasets. We provide the full set of prompts and corresponding generated audio samples in Section 1 of our webpage.

## *sarīr* Dataset

In order to train our GenAI models, we create a bespoke audio dataset that captures the characteristics of *sarīr*. The goal is to develop a dataset that is coherent, representing the recognizable sonic space of *sarīr*, while being diverse enough to reflect the variations that naturally occur within calligraphic practice.

The design of the dataset is informed by traditional techniques of Persian calligraphy in *nastaliq* (نستعلیق) style—the predominant calligraphy style in Iran. We focus on performing the fundamental gestural strokes that underlie *nastaliq* writing. These forms are not full letters or words, but they are basal reed movements used to construct them (see Figure 3). These gestures are especially suitable for the audio domain, as each motion produces a distinct sound event depending on its shape and length.

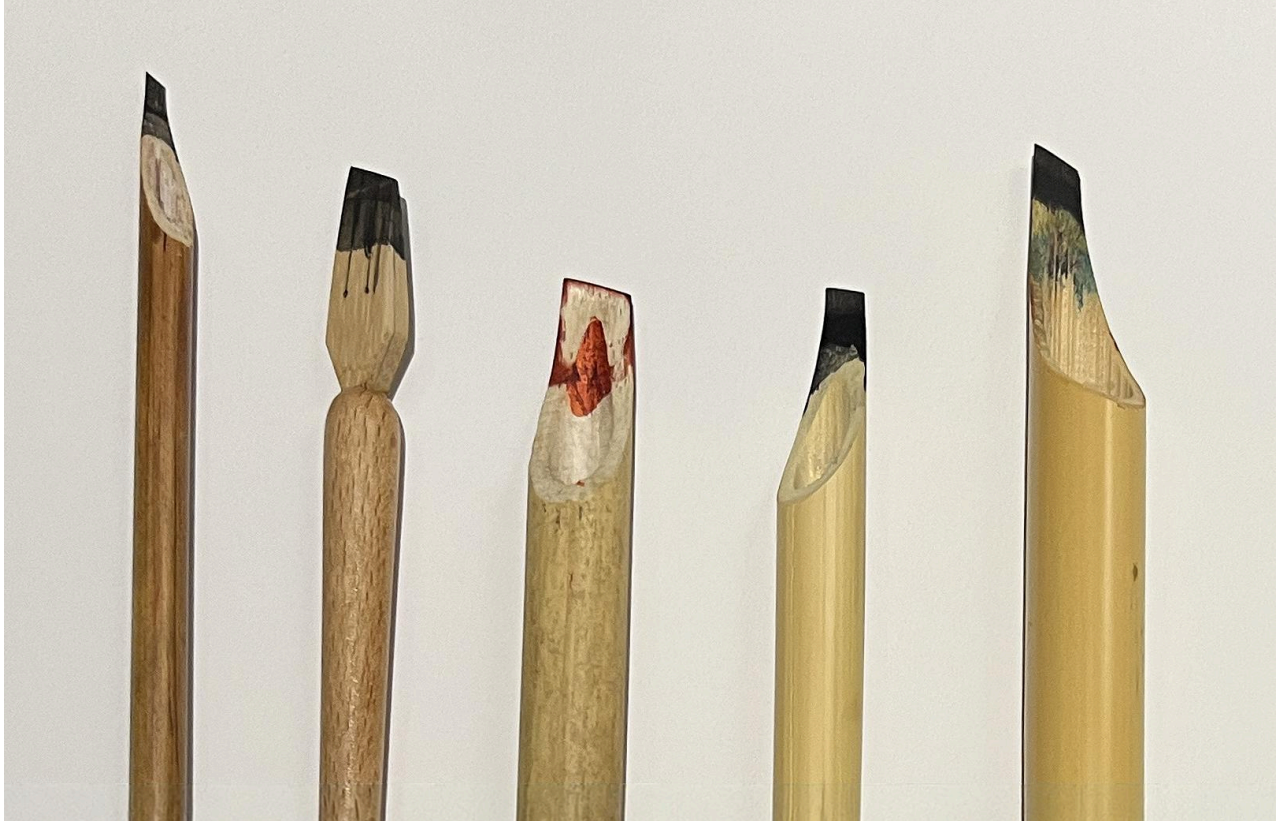

Figure 1. The selection of five calligraphy reed pens used for the dataset creation.

To ensure sonic diversity while maintaining aesthetic coherence, several variables were introduced during data creation:

- Reed pen size: multiple reed pens with different nib widths are used, as each produces distinct tonal textures and frequencies. Figure 1 shows the different reed pens used in the creation of the dataset.
- Hand pressure and speed: variations in pressure and writing speed are incorporated that influence the loudness, continuity, and granularity of the sound.
- Stroke form: a range of calligraphic strokes is performed, each generating unique sonic characteristics depending on curvature, length, and pressure. In addition to conventional strokes, the reed pen is also used in a more experimental manner, treating it as a sound-making instrument. This adds diversity while remaining within the recognizable sonic boundaries of *sarīr*. Figure 3 illustrates one of the reference sheets created during the process, where the strokes move gradually from traditional key gestures (right side) to more experimental movements (left side).

To capture the best possible audio fidelity and detail, we experiment with several recording setups, including using a dynamic cardioid microphone, a stereo field recorder, and a contact microphone (see Figure 2). Each setup produces distinct timbral characteristics, but after comparative listening and spectral analysis, the cardioid microphone is found to capture the most balanced and accurate representation of the sound, particularly in preserving its subtle high-frequency textures and low-frequency resonances. This setup is therefore used for the final dataset recordings.

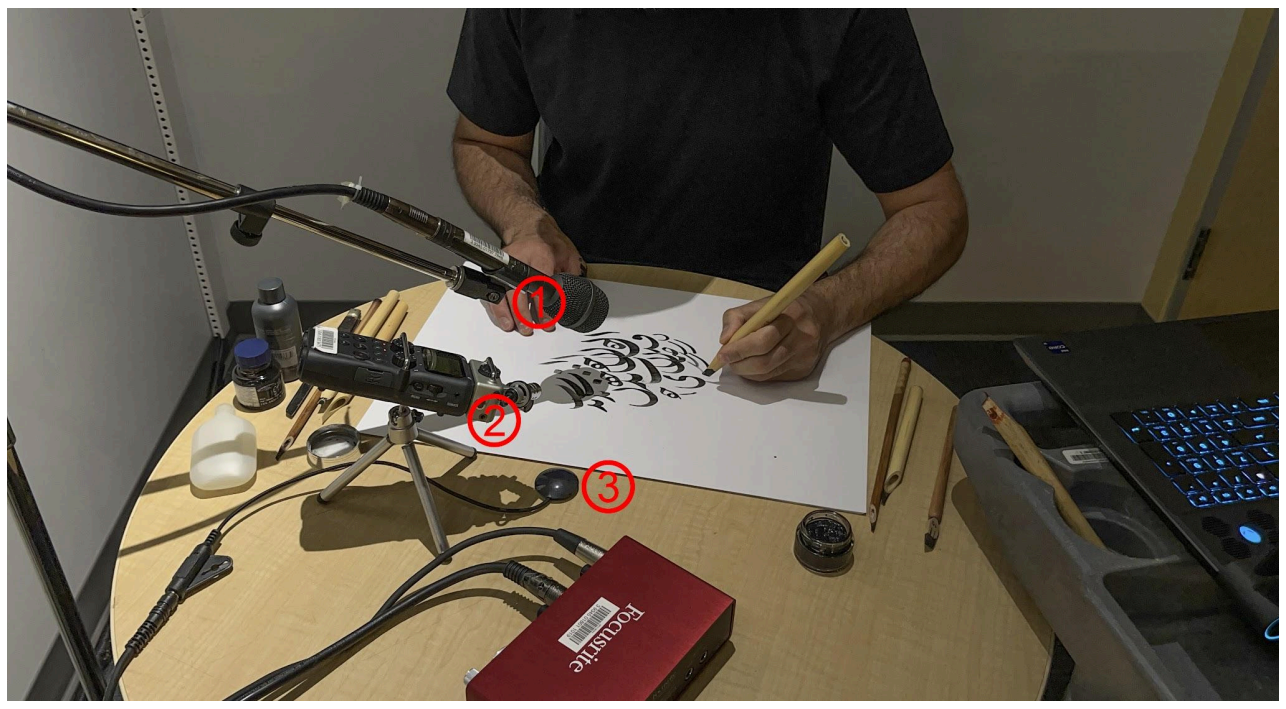

Figure 2. Recording setup using different microphones: 1) dynamic cardioid microphone. 2) stereo field recorder. 3) contact microphone.

In total, the dataset comprises 220 minutes of recorded material. Basic noise reduction and amplitude normalization are applied to the recordings to enhance clarity without compromising the natural texture and dynamic range of the sounds. The recordings are then analyzed to determine the average and maximum durations of individual sound events, since different calligraphic strokes produced sounds of varying lengths. We then cut the samples to standardized 6-second clips (the duration of the longest strokes), ensuring consistency across the dataset. Each audio clip contains a single instance of *sarīr* corresponding to one calligraphic gesture. The final dataset includes a total of 1787 individual audio samples. Examples from the dataset are presented in Section 2 of our webpage.

## Model Training

**Incorporated GenAI Architectures** To explore the opportunities of small data GenAI with our dataset, we incorporate two GenAI architectures: RAVE [35] and StyleGAN2 [36], both of which accommodate training on small datasets and have shown promising results in artistic use cases.

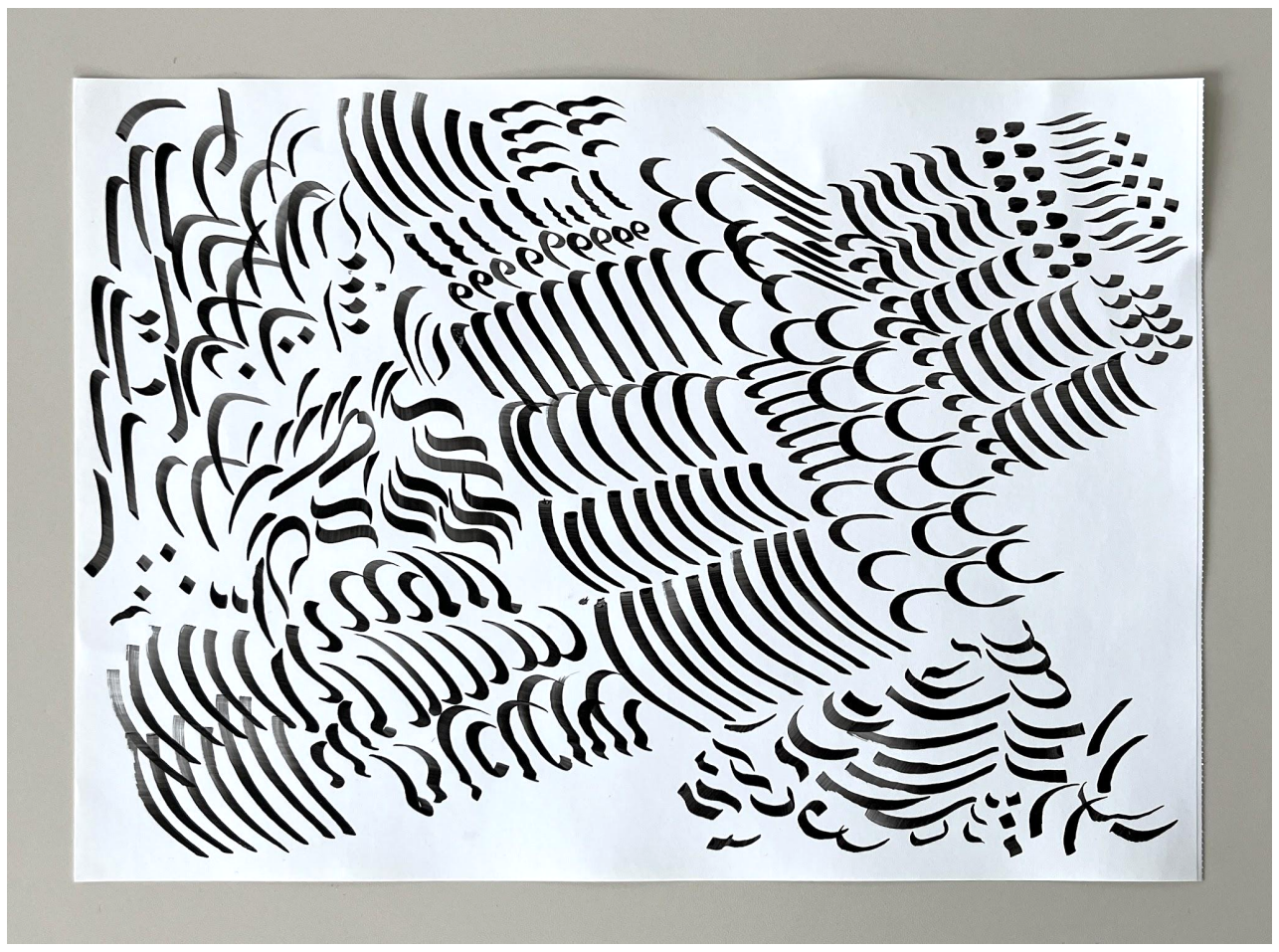

Figure 3. An example of a populated sheet during the dataset creation process.

RAVE is a neural audio synthesis model that learns a compact latent representation of the training data. Once trained on coherent and homogeneous data, it can generate new audio samples and provide an accessible latent space that allows direct manipulation of learned sonic features. RAVE also supports style transfer, mapping the perceptual characteristics of the training data onto external audio while preserving the input's temporal structure.

StyleGAN2, originally developed for image synthesis, can be repurposed for audio by generating spectrograms, which are visual time-frequency representations of sound (Figure 4). Its smooth latent space supports continuous interpolation between spectrograms, enabling the exploration of gradual transitions and novel variations. This property makes StyleGAN2 particularly compelling for creative audio experimentation based on spectral representations.

**Training Process** We train different RAVE configurations from the official repository [37] (v1, v2, v2_small, and v2+prior), which support different levels of audio quality and expressiveness. Models are trained with the recommended augmentation pipeline, and the best-performing checkpoints are chosen based on validation loss during the adversarial phase. Training is conducted on a single H100 GPU and requires approximately 11 hours per configuration.

To train StyleGAN2 on audio, each 6-second *sarīr* data sample is converted into a normalized 1024×1024 px spectrogram, following standard audio-analysis procedures. StyleGAN2 is then trained on two H100 GPUs for 800kimg with a batch size of 16. An initial run revealed mode collapse, which we addressed through a lightweight audio-level augmentation strategy by shifting silent regions and applying small random gain variations. This produced a stable training and a model capable of generating diverse samples. The final model is selected using the lowest validation FID score [6] (FID=9.37). Examples of the generated spectrograms can be seen in Figure 5, and more in Section 4.1 of our webpage.

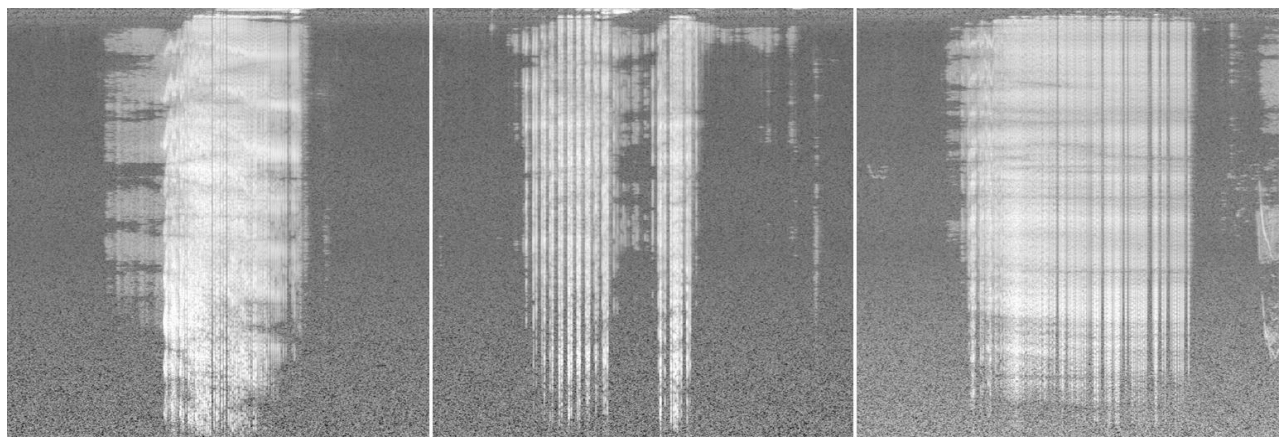

Figure 4. Examples of spectrograms created based on the *sarīr* audio samples.

## Results

**Generated Results from RAVE Models** Across the four trained RAVE configurations, we obtained models that captured the recognizable sonic signatures of *sarīr* with a high degree of clarity and accuracy. While all configurations produced coherent outputs, v2 and v2+prior models performed best comparatively, generating sounds that preserved both the fine-grained textures and broader gestural qualities of the dataset.

To explore the generative affordances of RAVE, we use the provided Max/MSP plugin[2], allowing real-time audio processing and control over the latent space. This environment offers an intuitive way to manipulate the learned sonic features and treat the model as a playable instrument. Interacting with the latent dimensions revealed meaningful structure, with variable dimensions corresponding to perceptually relevant features that make sense within the audio domain of *sarīr*. Multiple examples from the RAVE models are presented in Section 3 of our webpage.

Overall, the generated results demonstrate that the RAVE model can reliably learn the subtle timbral nuances of *sarīr*, producing outputs that preserve its fine textures, gestural dynamics, and recognizable auditory character presented in the dataset. The resulting model is not only effective for generating new sonic material but also locally accessible, directly manipulable, and fast enough to support real-time artistic experimentation.

**Generated Results from StyleGAN2** The StyleGAN2 model produces high-quality and diverse generated spectrograms (Figure 5). After reconstruction into audio files, these spectrograms yield new instances of *sarīr* that preserve its core sonic characteristics. The generated samples exhibit variations in timbre, texture, and duration while remaining aligned with the recognizable high-level qualities that define the *sarīr* sound.

After reconstruction, we observe a recurring artifact in the audio outputs: an unnatural timbral component not present in the original *sarīr* recordings. We attribute this issue to limitations inherent to spectrogram-based synthesis, particularly the loss of phase information during generation and reconstruction. To address this, we use our trained RAVE model as a post-processing filter: StyleGAN2-generated samples are fed into RAVE's style-transfer mode. Because the input already belongs to the same sonic domain that RAVE has learned, RAVE largely preserves the intended features of the signal while removing the unwanted artifact. This produces much cleaner results and restores the perceptual coherence of the generated sounds. Multiple examples of generated spectrograms, the reconstructed audios, and RAVE-corrected results are available in Section 4.1 of our webpage.

[2] Available at: https://forum.ircam.fr/projects/detail/nn/

Training StyleGAN2 on the *sarīr* dataset provides the ability to generate an unlimited number of new instances drawn from its latent space. Each generated spectrogram corresponds to a 6-second audio segment, matching the structure of the training data. Beyond these static samples, it also provides an expanded creative affordance by exploring StyleGAN2's capacity for smooth latent-space interpolations (see an example in Section 4.2 of our webpage). By continuously navigating the latent space and concatenating frames from the evolving spectrograms, we synthesize extended audio excerpts of desired length (see Figure 6).

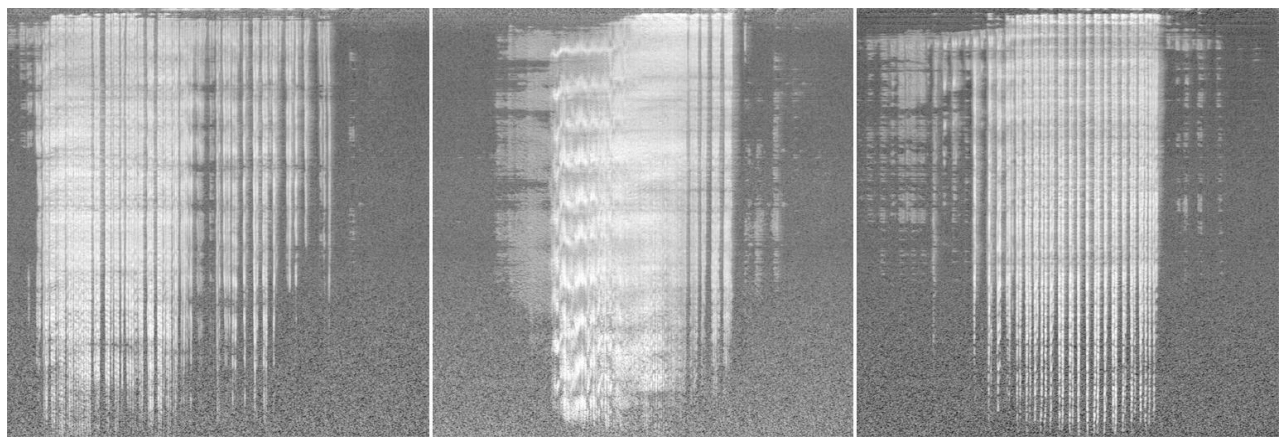

Figure 5. Examples of spectrograms generated from the trained StyleGAN2 model (available for listening on the results webpage).

This interpolation-based approach produces outputs that retain the core identity of *sarīr* while revealing new temporal trajectories shaped by movement through the latent space. The resulting sounds carry distinctive qualities not present in the original dataset, offering a novel and creative mode that merges the model's latent dynamics with the temporal logic of *sarīr* gestures. This opens significant room for artistic experimentation, both through direct manipulation of the latent space and through shaping the concatenation algorithm that assembles longer spectrograms. Examples of these concatenated samples are presented in Section 4.3 of our webpage.

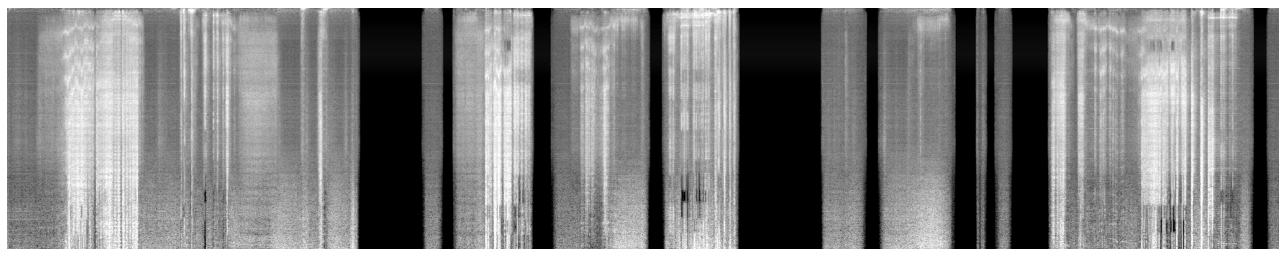

Figure 6. Example of an extended spectrogram generated from a custom concatenation of a latent interpolation using the StyleGAN2 model.

**Artistic Integration** The two trained models, based on RAVE and StyleGAN2, each introduce distinct creative affordances that expand the artistic possibilities of working with *sarīr*. RAVE offers high-fidelity, real-time audio generation and an immediately playable latent space, allowing fine-grained control over timbre, gesture, and modulation. The StyleGAN2 model enables a different mode of artistic engagement. The controlled generation of spectrogram interpolations and the algorithmic structuring of temporal evolution via long-form concatenations introduce expressive distortions, emergent textures, and novel sonic trajectories. Together, these affordances open generative pathways that are unavailable through large-scale text-to-audio systems.

We integrate both models into a series of artistic sketches that inform new artworks as well as updated versions of our earlier pieces and performances. Drawing from our prior work with a bespoke visual GAN trained on Persian calligraphy [38], we are developing an interconnected system in which the latent spaces of the audio and visual models influence one another directly. This multimodal ecosystem allows visual gestures to modulate sonic features and vice versa, forming a unified generative environment that can operate in both real-time and offline contexts. Such integration reimagines calligraphy as a dynamic audio-visual field rather than a solely visual artifact, foregrounding the multimodal nature of the practice as it is understood within its cultural context. Our artistic sketches are presented in Section 5 of our webpage.

## Discussion

Our findings show that a small data approach can generate high-quality and nuanced *sarīr* sounds that meaningfully capture the essence of the sonic domain, including its subtle timbral details and characteristic gestural qualities. In contrast, state-of-the-art large-scale text-to-audio models produce entirely unrelated outputs, demonstrating no awareness of the domain's cultural, material, or acoustic features. This stark disparity highlights how culturally specific artistic domains can fall completely outside the representational scope of large pretrained models, whose broad training corpora offer little to no exposure to such situated practices. These results illustrate that small-data GenAI methods offer a viable, and in some cases necessary, alternative for culturally specific creative work.

Reflecting on the process also illustrates how a small-data approach creates space for tacit, embodied, and culturally situated knowledge to shape the generative system. Key design decisions (such as determining what constitutes a single *sarīr* instance, identifying which variations enrich the dataset while keeping it coherent, and selecting augmentation methods that preserve the integrity of the sonic domain) are deeply informed by the domain knowledge and not arbitrary technical choices.

While some decisions inevitably carry elements of personal preference, many are grounded in the first author's embodied familiarity with the practice of calligraphy and the culturally informed understanding of what falls within, or falls outside, the subtle boundaries of the *sarīr* sound. These forms of knowledge allow the dataset to remain anchored to the defining features of the domain, demonstrating how cultural expertise becomes an active and necessary component in the process. This also illustrates how small-data approaches naturally support personalization, a benefit emphasized in prior research [14, 16].

Alongside this, the case study highlights the ongoing negotiation between cultural insight and the technical affordances of generative models. Many decisions arise at the intersection of what the models require and what the cultural material affords. For example, the need for square spectrograms for StyleGAN2 training informs the decision to regularize audio clips to the same length, which follows the stroke length diversity in the dataset, and also satisfies the RAVE data requirements. At the same time, culturally informed evaluation determines which preprocessing and augmentation strategies maintain the integrity of the sound and which would push it outside its meaningful range. These decisions are thus co-produced through artistic judgment and system-specific requirements and reveal how the small-data approach is shaped by a continual dialogue between cultural practice and technical design.

Despite its benefits, this approach also carries certain limitations and challenges. Whereas prompt-based systems offer low-effort interfaces and immediate results, albeit constrained by the need to verbalize artistic intent, small-data methods require deeper technical engagement. Even though RAVE and StyleGAN2 are among the popular, well-documented, and overall relatively accessible architectures, training and troubleshooting these models still demand technical skill, familiarity with machine learning concepts, and a willingness to work through implementation details. These challenges have motivated initiatives such as Autolume [25], which aim to provide non-technical, artist-oriented tools for small data GenAI. In our case study, we recognize that our ability to work at this depth is supported by our own technical literacy and experience, a form of privilege that not all artists or creatives may share. This further highlights the importance of continued efforts to make small-data approaches more accessible to a wider range of artists and cultural practitioners.

This focused case study also exemplifies the value of examining culturally specific material at a fine-grained level. By concentrating on the particular requirements of a narrowly defined domain, we are able to observe disparities in how different generative approaches respond to it, and to assess the effectiveness of small-data model crafting with much greater precision than broad, surface-level evaluations of cultural bias typically allow. Such focus enables a more granular analysis of cultural alignment that attends to the actual practices, materials, and aesthetics at stake, rather than abstract notions of representation. We therefore encourage further case studies, using a small-data approach, that work with concrete artistic processes and diverse cultural elements, treating dataset creation and model training as a site of cultural resistance to the biases and aesthetic flattening imposed by corporate large-scale GenAI.

## Conclusion and Future Works

This paper demonstrates how a small data approach enables culturally specific artistic creation by centring artist-curated datasets, domain expertise, and iterative model crafting. Through the case study of *sarīr*, we show how bespoke datasets and targeted training can capture the fine-grained aesthetic qualities of a culturally situated practice that lies far beyond the representational capacity of large-scale GenAI models. Our results highlight how the small data approach can preserve cultural nuance, support artistic control, and open expressive possibilities that emerge directly from situated knowledge and materially grounded processes.

We hope this work encourages further explorations of culturally specific GenAI across other artistic domains, especially those marginalized or misrepresented by mainstream models. Future work will involve expanding access to small-data methodologies by addressing barriers in technical literacy, tooling, and infrastructure for artists without machine learning expertise. We also see value in investigating hybrid approaches that combine small-data model crafting with fine-tuning of large-scale models, examining how these techniques might balance cultural specificity with broader generative capacities.

## Acknowledgmements

This research was supported by cloud computing resources provided by the Digital Research Alliance of Canada through the Metacreation Lab for Creative AI at Simon Fraser University.